# $f_B$ quenched and unquenched[*]

C. Bernard,[a] T. Blum,[b] A. De,[a] T. DeGrand,[c] C. DeTar,[d] Steven Gottlieb,[e] Urs M. Heller,[f] J. Hetrick,[b] N. Ishizuka,[a] J. Labrenz,[g] K. Rummukainen,[e] A. Soni,[h] R. Sugar,[i] D. Toussaint,[b] and M. Wingate[c]

[a]Department of Physics, Washington University, St. Louis, MO 63130, USA

[b]Department of Physics, University of Arizona, Tucson, AZ 85721, USA

[c]Physics Department, University of Colorado, Boulder, CO 80309, USA

[d]Physics Department, University of Utah, Salt Lake City, UT 84112, USA

[e]Department of Physics, Indiana University, Bloomington, IN 47405, USA

[f]SCRI, The Florida State University, Tallahassee, FL 32306-4052, USA

[g]Physics Department, University of Washington, Seattle, WA 98195, USA

[h]Physics Department, Brookhaven National Laboratory, Upton, NY 11973, USA

[i]Department of Physics, University of California, Santa Barbara, CA 93106, USA

Results for $f_B$, $f_{B_s}$, $f_D$, $f_{D_s}$, and their ratios are presented. High statistics quenched runs at $\beta = 5.7, 5.85, 6.0$, and 6.3, plus a run still in progress at $\beta = 6.52$ make possible a preliminary extrapolation to the continuum. The data allows good control of all systematic errors except for quenching, although not all of the error estimates have been finalized. Results from configurations which include effects of dynamical quarks show a significant deviation from the quenched results and make possible a crude estimate of the quenching error.

For the past two years, we have been computing heavy-light decay constants with Wilson fermions in the quenched approximation, and have more recently begun to use $N_F = 2$ dynamical staggered fermion background configurations to test directly the effects of quenching. Computations on the largest lattices have been performed on the 512-node and 1024-node Intel Paragon computers at Oak Ridge National Laboratory; Paragons at Indiana University and at the San Diego Supercomputer Center have been used for smaller lattices. Basic features of the calculation are described in Ref. [1]. Table 1 shows the parameters of the lattices used.

Quenched runs A, E, C, D, and H are on lattices with very nearly the same physical volume. Finite volume errors in the quenched approximation are determined by comparing the results of runs A and B. More work is needed to estimate the finite volume errors for the $N_F = 2$ dynamical fermion runs.

In most cases, we compute the pseudoscalar decay constant for static-light, as well as heavy-light, mesons. However, as explained in [1], the current procedure does not produce usable static-light results on lattices B, E, I and J. Efforts to remedy this situation are in progress.

For heavy-light mesons we use the Kronfeld-Mackenzie norm ($\sqrt{1-6\tilde\kappa}$) and adjust the measured meson pole mass upward by the difference between the heavy quark pole mass ("$m_1$") and the heavy quark dynamical mass ("$m_2$") as calculated in the tadpole-improved tree approximation [2]. Since we do not make the additional changes in action and operators called for by the heavy Wilson quark program [3], some $\mathcal{O}(ma)$ errors will still be present. Further $\mathcal{O}(a)$ errors are

---

[*]presented by C. Bernard at *Lattice '95*, Melbourne, Australia, July 11-15, 1995. To be published in the proceedings. Preprint Wash. U. HEP/95-31

2Table 1
Lattice parameters. Runs F, G, I, J, and K use variable-mass Wilson valence quarks and two flavors of fixed-mass staggered dynamical fermions; all other runs are quenched with Wilson valence quarks. The numbers in parenthesis are the planned final number of configurations for runs still in progress. F lattices were provided by the Columbia group; G, by HEMCGC

| name | $\beta$ | size | # configs. |
|---|---|---|---|
| A | 5.7 | $8^3 \times 48$ | 200 |
| B | 5.7 | $16^3 \times 48$ | 100 |
| E | 5.85 | $12^3 \times 48$ | 100 |
| C | 6.0 | $16^3 \times 48$ | 100 |
| D | 6.3 | $24^3 \times 80$ | 100 |
| H | 6.52 | $32^3 \times 100$ | 18 (50) |
| F | 5.7 $m = 0.01$ | $16^3 \times 32$ | 49 |
| G | 5.6 $m = 0.01$ | $16^3 \times 32$ | 95 (200) |
| I | 5.445 $m = 0.025$ | $16^3 \times 24$ | 118 |
| J | 5.47 $m = 0.05$ | $16^3 \times 24$ | 128 |
| K | 5.415 $m = 0.0125$ | $8^3 \times 24$ | 78 |

introduced by the Wilson light quark. An extrapolation to $a = 0$ is thus crucial.

Since we only have results for degenerate light quarks, we determine the strange quark hopping parameter $\kappa_s$ by adjusting the pseudoscalar mass to $\sqrt{2m_K^2 - m_\pi^2}$, the lowest order chiral perturbation theory value.

A plot of $f_P \sqrt{M_P}$ vs. $1/M_P$ is shown in fig. 1 for lattice D. The fits are to the form $c_0 + c_1/M_P + c_2/M_P^2$. Because the systematic errors are expected to be larger for the heavier propagating quarks, we use fits like that shown to the "lighter heavies" (with meson masses lighter than the D) on each lattice to give our central values for decay constants. Fits to the "heavier heavies" (with meson masses from the D region up to $\approx 4$ GeV) give one estimate of the systematic error caused by large lattice masses.

Figure 2 shows $f_B$ plotted versus lattice spacing, as determined from $f_\pi$. Extrapolation to $a = 0$ of the quenched data by a linear fit seems justified; the fit has $\chi^2$/d.o.f. = 1.2 and gives $f_B = 151(5)$MeV. We are unwilling at this point to attempt an extrapolation of the dynamical fermion data. The present data from the three such runs at the largest values of $a$ (runs I, J and K) is suspect since the lattices are short in the time direction ($N_t = 24$) and the plateaus we do have are rather poor.

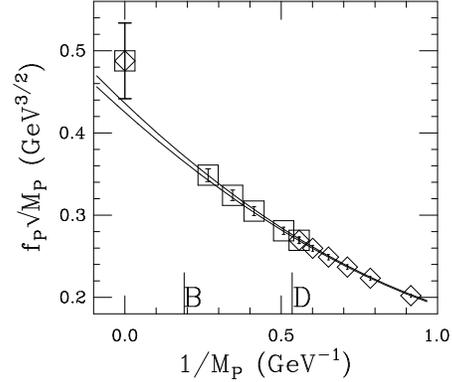

Figure 1. $f_P(M_P)^{\frac{1}{2}}$ vs. $1/M_P$ for lattice D. The lower line is a covariant fit ($\chi^2$/d.o.f. = 1.1) to the diamonds ("lighter heavies" + static); the slightly higher line is a covariant fit ($\chi^2$/d.o.f. = 2.7) to the squares ("heavier heavies" + static). The light quark is extrapolated to the physical mass $(m_u + m_d)/2$. The scale is set by $f_\pi$.

Work is in progress on longer versions of the lattices at large $a$ and on lattices at intermediate values of $a$. We hope that an extrapolation of the dynamical fermion data to the continuum will be possible when that work is completed.

The systematic errors are estimated by varying the parameters and fits used in the computation. Table 2 shows central values and error estimates for several decay constants and ratios from various versions of the analysis. Some errors are estimated in more than one way, and not all the different errors are in principle independent. The versions are as follows:

0) Central values: Fits to "lighter heavies" with $f_\pi$ scale and linear extrapolation in $a$. Boosted couplings approximately given by $g_{MF}^2 = g^2/$(plaquette) but with no tadpole improvement of the zero-mass axial renor-malization constant $Z_A$. Fields normalized by $\sqrt{1-6\tilde{\kappa}}$; meson mass shifted upward by $m_2 - m_1$; $\kappa_s$ fixed from the pseudoscalar mesons as described above.

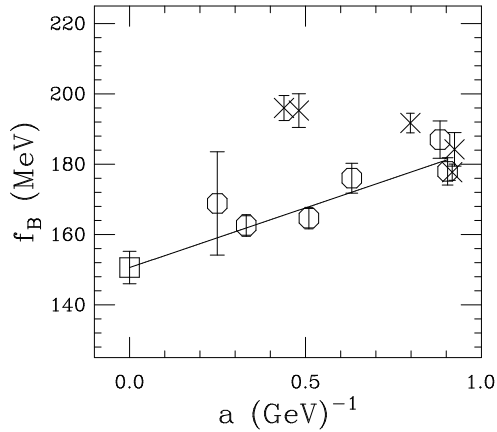

Figure 2. $f_B$ vs. $a$. Octagons are quenched lattices (runs A, B, E, C, D, H); crosses are $N_F = 2$ dynamical fermion lattices (runs F, G, I, J, K). A linear fit to the quenched points only is shown; the extrapolated value at $a = 0$ is indicated by the square. The slightly lower quenched result at $a \approx 0.9 \text{GeV}^{-1}$ comes from run B; the higher one, from run A. The scale is set by $f_\pi = 132$ MeV throughout.

1) Statistical errors in version 0. Experience shows that systematic errors due to choice of fitting ranges in $t$ of the correlators are comparable to the statistical errors.

2) Errors due to interpolation in $1/M$ at fixed $a$; estimated by replacing correlated $1/M$ fits in version 0 by uncorrelated fits.

3) $\mathcal{O}((ma)^2)$ errors: Replace "lighter heavies" in version 0 by "heavier heavies." ($\mathcal{O}(am)$ errors are presumably corrected for by the linear extrapolation in $a$.)

4) $\mathcal{O}((ma)^2)$ errors: Instead of $m_2 - m_1$, shift the meson mass by $m_3 - m_1$ where $m_3$ is the "hyperfine splitting" mass ($1/m_3$ is the coefficient of $\sigma \cdot B$) [2].

5) Finite volume errors: Take percent difference between runs A and B and apply that percentage to final, extrapolated values.

6) $\mathcal{O}(a^2)$ errors: Replace linear extrapolation in $a$ by linear plus quadratic extrapolation.

7) Weak coupling perturbation theory error: If boosted couplings in version 0 are expressed as $g_V^2(q^*)$ [4] then $q^* \approx 4.3/a$. Change $q^*$ to $2.2/a$ (still with no tadpole improvement of $Z_A$) and compare.

8) Error from fixing $\kappa_s$: Determine $\kappa_s$ by adjusting mass of vector $s\bar{s}$ state ($\phi$) to physical value and compare with version 0.

9) Error from fixing the scale: Use $m_\rho$ instead of $f_\pi$ to set the scale.

10) Quenching: Compare result from run G to the quenched results (à la version 0) interpolated to the same lattice spacing, and take same percentage error on final extrapolated result.

11) Quenching: Same as version 10, but also fix $\kappa_s$ from the $\phi$ everywhere, as in version 8.

Table 2
Central values and errors from various versions of the analysis for $f_B$, $f_{B_s}/f_B$, $f_{D_s}$, and $f_{D_s}/f_D$. See text for description of the versions. Decay constants and errors are in MeV.

| version | $f_B$ | $f_{B_s}/f_B$ | $f_{D_s}$ | $f_{D_s}/f_D$ |
|---|---|---|---|---|
| 0) | 151 | 1.11 | 198 | 1.09 |
| 1) | 5 | 0.02 | 5 | 0.01 |
| 2) | 10 | 0.00 | 3 | 0.01 |
| 3) | 1 | 0.02 | 2 | 0.00 |
| 4) | 5 | 0.00 | 3 | 0.01 |
| 5) | 8 | 0.01 | 4 | 0.01 |
| 6) | 4 | 0.01 | 6 | 0.03 |
| 7) | 5 | 0.01 | 3 | 0.00 |
| 8) | – | 0.07 | 7 | 0.05 |
| 9) | 11 | 0.08 | 1 | 0.04 |
| 10) | 26 | 0.00 | 14 | 0.01 |
| 11) | – | 0.03 | 19 | 0.02 |

The systematic error within the quenched approximation is now determined by adding in quadrature the errors determined by versions 1 (used as estimate of errors in $t$ fits) + 2 + 5 + 6 + 7 + the larger of 3 or 4. Note that if version 6 were guaranteed to be a good estimator of the error in the extrapolation to the continuum, versions 3 or 4 would be superfluous. However, since the error in these error estimates is probably comparable to the errors themselves, we take a more conservative approach and include the larger of versions 3 or 4 in addition to version 6.

Once the quenched approximation computation has been reliably extrapolated to the continuum, any dependence of the results on how the scale is fixed (version 9) or how $\kappa_s$ is fixed (version 8) must be considered as an error due to quenching itself. Assuming our extrapolation is reliable, we therefore estimate the quenching error by taking the largest of the errors determined by versions 8, 9, 10, or 11.

It is clear that some of the estimates given in Table 2 are rather crude; all should be taken as preliminary. First of all, the use of perturbation theory should be rationalized by using tadpole improvement throughout and then comparing several reasonable choices for the scale of the coupling constant. Further, the effect of changing the fitting ranges in $t$ should be determined directly. Errors in the interpolation in $1/M$ are more naturally determined by changing the fitting function rather than the type of fit. C. Allton has suggested [7] estimating the large $am$ errors by comparing to an "old-style" analysis with $\sqrt{2\kappa}$ normalization and no mass shifts. He has further advocated estimating the error in the continuum extrapolation by fitting the three points with smallest $a$ to a constant. These approaches will be tried.

Finally, the quenching error can be more reliably estimated when the $N_F = 2$ dynamical fermion results can be extrapolated to the continuum. Note from fig. 2 that the error determined thereby may be considerably larger than the current estimate at fixed lattice spacing. Of course, at that point one may wish to take as central values the $N_F = 2$ results. However, there would still be a quenching error since one is (at least at present) only extrapolating the valence and not the dynamical quarks to the chiral limit. (This is a "partially quenched theory" [5].) The real world, moreover, has $N_F = 3$. In fact the $N_F = 2$ simulations, even after extrapolation to the continuum, may be just as far from the real world as those of the quenched approximation[6].

The (still preliminary) results are:

$$f_B = 151(5)(16)(26) \qquad f_D = 182(3)(9)(22)$$
$$f_{B_s} = 169(7)(14)(29) \qquad f_{D_s} = 198(5)(10)(19)$$
$$\frac{f_{B_s}}{f_B} = 1.11(2)(4)(8) \qquad \frac{f_{D_s}}{f_D} = 1.09(1)(4)(5)$$

where the first error is statistical; the second, the systematic error within the quenched approximation; the third, the estimate of the quenching error. Decay constants are in MeV. While the estimate of the quenching error has large uncertainties at this stage, the error within the quenched approximation will, we think, be determined rather well once the afore-mentioned refinements in the analysis are completed.

We thank C. Allton and S. Sharpe for useful conversations. Computing was done at ORNL Center for Computational Sciences, Indiana University, and SDSC. We thank the Columbia group and the HEMCGC collaboration for providing dynamical fermion configurations. This work was supported in part by the DOE and NSF.